\documentclass[12pt]{article}
\usepackage[T1]{fontenc}

\makeatletter

\providecommand{\LyX}{L\kern-.1667em\lower.25em\hbox{Y}\kern-.125emX\@}

\makeatother

\begin{document}

\baselineskip .3in \begin{titlepage}

{\par\centering \textbf{\large Market application of the percolation model: Relative
price distribution} \par}

{\par\centering \vskip .2in \par}

{\par\centering \textbf{Anirban Chakraborti}\( ^{(a),\: (b),\: } \)\footnote{%
\emph{email address} : anirban@cmp.saha.ernet.in
} \par}

{\par\centering \textit{\( ^{(a)} \)Saha Institute of Nuclear Physics}, \textit{1/AF
Bidhan Nagar, }\par}

{\par\centering \textit{Kolkata-700 064, India.}\par}

{\par\centering \emph{\( ^{(b)} \)Institute for Theoretical Physics, Cologne
University, }\par}

{\par\centering \emph{50923 Cologne, Germany}\par}

\noindent \vskip .3in 

\noindent \textbf{Abstract} 

\noindent We study a variant of the Cont-Bouchaud model which utilizes the percolation
approach of multi-agent simulations of the stock market fluctuations. Here,
instead of considering the relative price change as the difference of the total
demand and total supply, we consider the relative price change to be proportional
to the ``relative'' difference of demand and supply (the ratio of the difference
in total demand and total supply to the sum of the total demand and total supply).
We then study the probability distribution of the price changes.

\vskip 0.5in


\noindent \end{titlepage}

\noindent \newpage

\section{Introduction}

Statistical physics contains the methods for extracting the average properties
of a macroscopic system (matter in bulk) from the microscopic dynamics of the
systems. It also gives us precise knowledge of the fluctuations (above these
averages) of these quantities \cite{Stat}. Scaling laws, experimental or theoretical,
have been of special interests to physicists. Hence, physicists are trying to
employ these methods to study the fluctuations of the stock markets (the study
of which began with the work of Louis Bachelier in 1900 \cite{LB}) as well.
With access to large sets of data from financial markets, an extensive search
for such scaling laws has begun recently \cite{Ecophy}. Fluctuations over the
average, say in some stock prices, are of immense interest to the economists
also. The nature of these fluctuations, whether random or otherwise, are of
extreme importance. Stigler \cite{Stig} studied the market fluctuations by
employing Monte Carlo methods more than thirty years back. The fluctuations
are believed to follow a Gaussian distribution for long time intervals. Mandelbrot
\cite{Man} was first to observe a clear departure from Gaussian behaviour for
these fluctuations for short time intervals. There have been various explanations
and descriptions for it, ranging from power laws, exponentials to multi-fractal
behaviour.

The percolation \cite{SAS} approach of Cont and Bouchaud \cite{CB} is one
of the simplest of the numerous multi-agent simulations of the stock market
fluctuations. Monte Carlo simulations of the above model made at the percolation
threshold, show power-law ``fat'' tails for short time intervals and exponential
truncation for longer time intervals. The model is also consistent with the
weak correlations between successive changes of price and strong correlations
between successive values of changes of price. There has been various variants
of the Cont-Bouchaud model \cite{ECB}. Here, we study another variant of the
Cont-Bouchaud model where instead of considering the relative price change coming
from the difference of the total demand and total supply, we consider the relative
price change to be proportional to the ``relative'' difference of demand and
supply (i.e., the ratio of the difference in demand and supply to the sum of
demand and supply).

\section{The model and results}

We human beings are ``social animals'' and hence like to stay together and
are influenced by others at all spheres of life. At most occasions, we form
``clusters'', which for simplicity, will be considered as random. As in the
case of percolation theory of random graphs, the traders are assumed to just
form random clusters and share their opinions. The history of the price changes
and the limitations in the disposable capital of each trader are ignored. 

The original Cont-Bouchaud model \cite{CB} considered the mean-field limit
of infinite-range interactions instead of the usual nearest-neighbour percolation
on lattices \cite{SAS}. In the variant models \cite{ECB}, the sites of a \( d \)-dimensional
lattice are randomly occupied with probability \( p \) and empty with probability
(\( 1-p \)) and the occupied nearest-neighbours form clusters. Each cluster
containing \( N_{t} \) traders decides randomly, to buy (with probability \( a \)),
sell (also with the same probability \( a \)), or to remain inactive (with
probability \( 1-2a \)). So far, the relative change of the price was considered
to be proportional to the difference between the total demand and the total
supply. Hence, for any time step \( \Delta t \), we first find the existing
clusters and the number \( n_{s} \) of clusters, each containing \( s \) traders.
Then each cluster randomly decides whether to buy, sell or remain inactive with
the above mentioned probabilities. The parameter \( a \) is called the ``activity''
and the increase in activity is equivalent to the increase in the time unit,
since \( a \) is the fraction of traders which are active per unit time. Thus
small \( a \) correspond to small time intervals and large \( a \) (with the
maximum of \( 0.5 \)) correspond to large time intervals. Then, the relative
price change for one time step is considered proportional to the difference
of the total demand and total supply:
\begin{equation}
\label{pr1}
R(t)=\ln P(t+\Delta t)-\ln P(t)\propto \sum _{s}n_{s}^{buy}s-\sum _{s}n_{s}^{sell}s
\end{equation}

\noindent where the constant of proportionality is taken to be unity. 

If we take one time step \( \Delta t \) to be very small so that only one cluster
of traders can trade during this time interval (the number of clusters trading
in one time step \( N=aN_{t}\sim 1 \)), then the probability distribution \( P(R) \)
is completely symmetric about zero (as in real stock markets) and just follows
the distribution \( n_{s} \) of clusters. The distribution, right at \( p=p_{c} \)
is \( n_{s}\propto 1/s^{\tau } \) with \( 2<\tau <2.5 \) in two to infinite
dimensions \cite{SAS}. If the time step \( \Delta t \) is large so that all
the traders can trade in each time step (\( N\sim N_{t} \)), then the probability
distribution \( P(R) \) is closer to a Gaussian. When the time step is in the
intermediate range so that \( 1\ll N\ll N_{t} \) the price changes are bell-shaped
with power-law tails. This crossover to Gaussian behaviour with the variation
of \( a \) is observed in reality also \cite{SK}.

In our model, the relative price change is proportional to the ``relative''
difference of demand and supply, i.e., the ratio of the difference in demand
and supply, and the grand total of demand and supply:
\begin{equation}
\label{pr2}
R(t)=\ln P(t+\Delta t)-\ln P(t)\propto \left( \sum _{s}n_{s}^{buy}s-\sum _{s}n_{s}^{sell}s\right) /\left( \sum _{s}n_{s}^{buy}s+\sum _{s}n_{s}^{sell}s\right) 
\end{equation}

\bigskip{}
\noindent where the constant of proportionality is again taken to be unity.

Our computer simulations first distribute sites randomly on the square lattice
of dimensions \( L\times L \) at the percolation threshold (\( p=p_{c}=0.592746 \))
and then determine the clusters. For each time step \( \Delta t \), we allow
each cluster to decide randomly whether to trade or remain inactive. The trading
clusters then again randomly decide to buy or sell, and then equation (\ref{pr1})
or equation (\ref{pr2}) determine the relative price change. We average over
many lattice configurations to find the probability distribution \( P(R) \).
Programs in C and FORTRAN, based on the Hoshen-Kopelman algorithm in two dimensions
(where we have considered site percolation with free boundary conditions) are
available from the author.

The histograms of price changes which we get when the relative price changes
are determined according to equation (\ref{pr1}) and those according to equation
(\ref{pr2}) are shown in Fig. 1, with \( p_{c}=0.592746 \). In the Cont-Bouchaud
model, we also see a crossover from a power-law to a bell-shaped behaviour (within
the accuracy of the computer simulations) for increase in activity \( a \)
(not shown in the figure) showing its similarity with real stock markets. In
this model, since the magnitude of the relative price change always lies between
zero and unity, we observe a sharp cut-off in the histogram, unlike in real
markets. Thus the original Cont-Bouchaud model is superior to this model, in
this respect.

\section{Discussions and Summary}

We study a variant of the Cont-Bouchaud model: the relative price changes are
defined as the ratio of the difference in demand and supply to the sum of demand
and supply, \( R(t)=\left( \sum _{s}n_{s}^{buy}s-\sum _{s}n_{s}^{sell}s\right) /\left( \sum _{s}n_{s}^{buy}s+\sum _{s}n_{s}^{sell}s\right)  \),
where the constant of proportionality is taken to be unity. We also present
some of the previous results of a variant of the Cont-Bouchaud model for comparison.
We observe a sharp cut-off in the histogram for this model, unlike in real markets,
which shows that the original Cont-Bouchaud model is superior to this model
in this respect. This model too could be made more realistic, e.g., including
the history of the price changes. 

\vskip 0.3in

\noindent \textbf{Acknowledgements }
\medskip{}

The author is grateful to D. Stauffer for his warm hospitality and useful discussions,
and the Graduate College of Scientific Computing, Cologne University, for partial
financial support.

\vskip 0.3in

\noindent \textbf{Figure captions} 

\vskip 0.2in

\noindent \textbf{Fig. 1 :} Histogram of relative price changes plotted in the
linear-logarithmic scale, obtained from computer simulations made at the percolation
threshold for \( 5000 \) square lattices of size \( 1001\times 1001 \), \( 1000 \)
time intervals and activity \( 0.01 \).

\end{document}